\documentclass[aps,prd,preprint,superscriptaddress,showpacs,floatfix,
preprintnumbers,nofootinbib,a4paper]{revtex4-1}
\usepackage{hyperref}
\usepackage{color}
\usepackage{graphicx}
\usepackage{textcomp}
\usepackage{bbm}
\usepackage{amsmath,amssymb}
\usepackage{enumerate}
\newcommand{\mathsym}[1]{{}}

\newcommand{\ba}{\begin{array}}
\newcommand{\ea}{\end{array}}
\newcommand{\be}{\begin{equation}}
\newcommand{\ee}{\end{equation}}
\newcommand{\beqa}{\begin{eqnarray}}
\newcommand{\eeqa}{\end{eqnarray}}

\begin{document}
\vspace*{1cm}
\title{Neutrino masses and mixing from flavour antisymmetry}
\bigskip
\author{Anjan S. Joshipura}
\email{anjan@prl.res.in}
\affiliation{Physical Research Laboratory, Navarangpura, Ahmedabad 380 009, India.}
%\pacs{11.30.Hv, 14.60.Pq, 11.30.Er}
%--------------------------------------------------------------
\begin{abstract}
\vspace*{0.2cm}
We discuss consequences of assuming ($i$) that the (Majorana) neutrino mass
matrix $M_\nu$ displays flavour antisymmetry, $S_\nu^T M_\nu S_\nu=-M_\nu$
with respect to some discrete symmetry $S_\nu$ contained in $SU(3)$ and
($ii$) $S_\nu$ together with a symmetry $T_l$ of the  Hermitian combination
$M_lM_l^\dagger$ of the charged lepton mass matrix forms a finite discrete
subgroup $G_f$  of $SU(3)$ whose breaking generates these symmetries.
Assumption ($i$) leads to at least one  massless neutrino and allows only
four textures for the neutrino mass matrix in a basis with   a diagonal
$S_\nu$ if it is assumed that the other two neutrinos are  massive. Two of
these textures  contain a degenerate pair of neutrinos.Assumption ($ii$) can
be used to determine the neutrino mixing patterns. We work out these
patterns for two major group series $\Delta(3 N^2)$ and $\Delta(6 N^2)$  as
$G_f$. It is found that all $\Delta(6 N^2)$ and $\Delta(3 N^2)$ groups with
even $N$ contain some elements which can provide appropriate $S_\nu$. Mixing
patterns can be determined  analytically for these groups and it is found
that only one of the four allowed neutrino mass textures is consistent with
the observed  values of the mixing angles $\theta_{13}$ and $\theta_{23}$.
This texture corresponds to one massless and a degenerate pair of neutrinos
which can provide the solar pair in the presence of some perturbations. 
The well-known groups $A_4$ and $S_4$ provide examples of the groups in
respective series allowing correct $\theta_{13}$ and $\theta_{23}$. An
explicit example based on $A_4$ and displaying a massless and two quasi
degenerate neutrinos is discussed.
\end{abstract}
%----------------------------------------------------------

\maketitle
\section{Introduction}
Orderly pattern of neutrino mixing appears to hide some symmetry, discrete
or continuous.  It is possible to connect a given mixing  pattern with some discrete
symmetries of the leptonic mass matrices.  Such symmetries may however be 
residual symmetries arising from a bigger  symmetry in the underlying theory.
 One can obtain a possible
larger picture by assuming that these symmetries are a part of a bigger
group operating at the fundamental level whose breaking leads to the
symmetries of the mass matrices. There is an extensive literature
on study of possible residual symmetries of the mass matrices and of the
groups which harbor them
\cite{Lam:2008rs,Lam:2008sh,Lam:2009hn,Lam:2011ag,Toorop:2011jn,
deAdelhartToorop:2011re,Altarelli:2012ss,Holthausen:2012wt,
Hu:2012ei,Hernandez:2012ra,Hernandez:2012sk,
Holthausen:2013vba,Holthausen:2013vba,Lavoura:2014kwa,Fonseca:2014lfa,
Hu:2014kca}, see
\cite{Altarelli:2010gt,King:2013eh,Smirnov:2011jv} for
reviews and additional references. 

Starting point in these approaches is to assume the existence
of some symmetries $S_\nu$ (usually a $Z_2\times Z_2$) and $T_l$ (usually
$Z_N,N\geq 3$) of the (Majorana) neutrino and the charged lepton mass
matrices
\beqa
T_l^\dagger M_lM_l^\dagger T_l=M_lM_l^\dagger~,\\
S_\nu ^T M_\nu S_\nu =M_\nu~.\eeqa
Matrices diagonalizing the  $3\times 3$ symmetry matrices $S_\nu,T_l$ can be
related to the mixing matrices in each sector. The structures
of these matrices  can also be independently fixed if one assume that $S_\nu$
and $T_l$
represent specific elements of some discrete group $G_f$ in a
given three dimensional representation. In this way, the leptonic mixing
can be directly related to group theoretical structures. This
reasoning has been  used for the determination of the
neutrino  mixing angles in case of the  three non-degenerate neutrinos
\cite{Lam:2008rs,Lam:2008sh,Lam:2009hn,Lam:2011ag,Toorop:2011jn,
deAdelhartToorop:2011re,Altarelli:2012ss,Holthausen:2012wt,
Hu:2012ei,Hernandez:2012ra,Hernandez:2012sk,
Holthausen:2013vba,Holthausen:2013vba,Lavoura:2014kwa,Fonseca:2014lfa,
Hu:2014kca},
two or three degenerate
neutrinos \cite{Hernandez:2013vya,Joshipura:2014qaa} and one massless and two
non-degenerate neutrinos \cite{Joshipura:2013pga,Joshipura:2014pqa}.  

The residual symmetries may arise from spontaneous breaking of $G_f$ 
if the vacuum expectation values of the 
Higgs fields responsible for generating leptonic masses break $G_f$ but
respect $S_\nu,T_l$. We wish to study in this paper consequences of an
alternative assumption that the spontaneous breaking of $G_f$
leads to an
$M_\nu$ which displays antisymmetry instead of symmetry, i.e. assume that
eq.(2) gets replaced by
\be 
S_\nu^T M_\nu S_\nu=-M_\nu~\ee
but (1) remains as it is. These assumptions prove to be quite powerful and
are able to simultaneously restrict both the mass patterns and mixing angles
when embedding of $S_\nu,T_l$ into $G_f$ is considered. We shall further
assume
that $S_\nu,T_l$ belong to some  finite discrete subgroup of $SU(3)$ with
Det$(S_\nu,T_l)=\pm 1.$ Then the  first consequence of imposing eq.(3) is
that
Det$M_\nu=0$, i.e. at least one of the neutrinos remains massless. Since
cases with two (or three !) massless neutrinos are not phenomenologically
interesting,  we shall restrict ourselves to cases with only one massless
neutrino. Then as a second consequence of eq.(3), one can determine all the
allowed forms of $M_\nu$ in a given basis for all possible $S_\nu$
contained in $SU(3)$. There exist only
four possible $M_\nu$ (and their permutations) consistent with eq.(3) in a
particular basis
with a diagonal $S_\nu$. Two of these give one massless and two
non-degenerate
neutrinos and the other two  give a massless and a degenerate pair of
neutrinos which may be identified with the solar pair.

We determine all the allowed textures of the  neutrino mass matrix in the
next
section. Subsequently, we discuss groups $\Delta(3N^2)$ and $\Delta(6N^2)$ 
and identify those which  can give correct
description of mixing using flavour antisymmetry.  In section IV, we
introduce
$Z_2\times Z_2$ as neutrino residual symmetry and present an example in which
neutrino mass matrix gets fully determined group theoretically except for
an overall scale. We discuss a realization of the  basic idea
with a simple example based on the $A_4$ group in section V. Section VI
contains summary and comparison with earlier relevant works. 

\section{Allowed textures for neutrino mass matrix}
We shall first consider the case of only one $S_\nu$ satisfying eq.(3) and
subsequently generalize it to include two. The unitary matrix 
$S_\nu$ can be diagonalized by another unitary
matrix $V_{S_\nu}$:
$$V_{S_\nu}^\dagger S_\nu V_{S_\nu}=\tilde{S}_\nu$$
where $\tilde{S}_\nu$ is a diagonal matrix having the form:
\be \label{ds}
\tilde{S}_\nu={\rm diag.} (\lambda_1,\lambda_2,\lambda_3)~,\ee
Unitarity of $S_\nu$ implies that $\lambda_{1,2,3 }$ are some 
roots of unity. They are  related by the
condition $Det S_\nu=+1$ which we assume without lose of generality.
We now go to the basis with a diagonal $S_\nu$. Defining
$\tilde{M}_\nu=V_{S_\nu}^TM_\nu V_{S_\nu}$, eq.(3) can be rewritten as:
\be\label{con1}
(\tilde{M}_\nu)_{ij}(1+\lambda_i \lambda_j)=0~~~~~(i,j~~{\rm
not~~summed)}~.\ee
It follows that a given element $(\tilde{M}_\nu)_{ij}$ is non-zero
only
if
the factor in bracket multiplying it is zero. This cannot happen for an
arbitrary set of $\lambda_i$ and one needs to impose specific relation
among them to obtain a non-trivial $\tilde{M}_\nu$. We now argue that
only
two possible forms of $\tilde{S}_\nu$ and their permutations  lead to 
neutrino mass matrices
with two massive neutrinos. The third mass will always be zero as a
consequence of eq.(3) and the assumption that $S_\nu$ belongs to 
$SU(3)$. These forms of $\tilde{S}_\nu$ are given by:
\beqa \label{dsforms}
\tilde{S_1}_\nu&=&{\rm diag.} (\lambda,-\lambda^*,-1)~,\nonumber \\
\tilde{S_2}_\nu&=&{\rm diag.}(\pm i,\mp i,1)~.\eeqa
$\lambda$ is an arbitrary root of unity.
This can be argued as follows. Assume that at least one off-diagonal element
of $\tilde{M}_\nu$ is non-zero which we take as the 12 element for
definiteness.
In this case, eq.(\ref{con1}) immediately
implies the first of eq.(\ref{dsforms}) as a necessary condition. One
can distinguish 
three separate cases of this condition\footnote{$\lambda=-1$ case
corresponds to permutation of the case with $\lambda=1$. } 
(I) $\lambda=1$ (II) $\lambda=\pm i $ and (III) $\lambda \not=\pm 1,\pm i$. The
entire
structures of $\tilde{M}_\nu$ get determined in these cases from 
condition eq.(\ref{con1}) as follows:
\be \label{text1}
\ba{ccc}
{\rm Texture~ I:~}&\tilde{S_1}_\nu=(1,-1,-1);~~&~~
\tilde{M}_\nu=m_0\left(\ba{ccc} 0&c&s e^{i\beta}\\
c&0&0\\
se^{i\beta}&0&0\\
\ea\right)\\
\ea
~,\ee
where $c=\cos\theta,s=\sin\theta$. This structure implies one massless and
two degenerate neutrinos with a mass $|m_0|$. In case of (II),
\be \label{text2}
\ba{ccc}
{\rm Texture~ II:~}&\tilde{S_1}_\nu=(\pm i,\pm i,-1);~~&~~
\tilde{M}_\nu=\left(
\ba{ccc} x_1&y&0\\
y&x_2&0\\
0&0&0\\
\ea\right)\\
\ea
~.\ee
This case corresponds to one massless and two non-degenerate neutrinos. In
the third case one gets
\be \label{text3}
\ba{ccc}
{\rm Texture~ III:~}&\tilde{S_1}_\nu=(\lambda,-\lambda^*,-1);~~&~~
\tilde{M}_\nu=m_0\left(
\ba{ccc} 0&1&0\\
1&0&0\\
0&0&0\\
\ea\right)~,\\
\ea
~~~~~ (\lambda\not=\pm 1,\pm i)\ee
which implies a massless and a pair of degenerate neutrinos. 

The cases (I,III) lead to the same mass spectrum but different mixing
patterns. $\tilde{M}_\nu$ in eq.(\ref{text1}) is diagonalized 
as $V_\nu^T \tilde{M}_\nu V_\nu={\rm diag.}(m_0,m_0,0)$ with
\be \label{unu1}
V_\nu=\left( \ba{ccc}
\frac{1}{\sqrt{2}}&-\frac{i}{\sqrt{2}}&0\\
\frac{c}{\sqrt{2}}&\frac{i c}{\sqrt{2}}&-s\\
\frac{s}{\sqrt{2}}e^{-i\beta}&\frac{i s}{\sqrt{2}}e^{-i\beta}&ce^{-i\beta}\\
\ea\right)
\left( 
\ba{ccc}
\cos\phi&-\sin\phi&0\\
\sin\phi&\cos\phi&0\\
0&0&1\\
\ea \right)
~.\ee
The arbitrary rotation by an angle $\phi$ originates due to degeneracy in
masses.
The texture II, eq.(\ref{text2}) is diagonalized by a unitary rotation in
the
12 plane while the one in  eq.(\ref{text3}) by a similar matrix with the
angle
$\frac{\pi}{4}$.

The permutations of entries in  $\tilde{M}_\nu$ give equivalent
structures
and 
are obtained by permuting entries in $\tilde{S_1}_\nu$. The case which is not
equivalent
to 
above textures  follows with a starting assumption that one of the
diagonal
elements of $\tilde{M}_\nu\not=0$ say,
$(\tilde{M}_\nu)_{11}\not=0$.
In this case one requires $\tilde{S}_\nu={\rm diag.}(\pm
i,\lambda^\prime,\mp
i\lambda^{*\prime})$ with $|\lambda^\prime|=1$.
The case with $\lambda^\prime=\pm i$ gives $\tilde{S_1}_\nu$ which is
already
covered.
$\lambda^\prime=\mp i$ implies the condition $\tilde{S_2}_\nu$ in
(\ref{con1}). This leads
to a new texture
\be \label{text4}
\ba{ccc}
{\rm Texture~ IV:~}&\tilde{S}_\nu=(i,-i,1);~~&~~
\tilde{M}_\nu=\left(
\ba{ccc} x_1&0&0\\
0&x_2&0\\
0&0&0\\
\ea\right)\\
\ea
~.\ee
For $\lambda^\prime=\pm 1$ one gets permutation of $\tilde{S_1}_\nu$ or
$\tilde{S_2}_\nu$ and
for $\lambda^\prime\not=\pm 1,\pm i$  only 11
element of $\tilde{M}_\nu$ is non zero and 
two neutrinos remain massless. Thus conditions
eq.(\ref{dsforms}) and their permutations exhaust all possible  textures of
$\tilde{M}_\nu$ consistent with the antisymmetry of $M_\nu$, eq.(3) 
and
two massive neutrinos. Any $G_f$ admitting an element with these sets of
eigenvalues will give  a viable choice for flavour antisymmetry group. Note
that texture III (IV) can be obtained from I(II) by putting $s(y)$ to zero.
But the residual symmetries in all four cases are different. Because of
this, the embedding groups $G_f$ can also be different. We therefore
discuss all these cases separately.

The mixing matrix in texture I contains two unknowns $\theta$ and $\beta$
apart from an overall complex scale $m_0$. This is a reflection of the fact
that the corresponding $S_\nu$ is a $Z_2$ symmetry and contains two
degenerate eigenvalues $-1$. These unknown can be fixed by imposing another 
residual $Z_2$ symmetry commuting with $S_\nu$ and satisfying eq. (2) or (3). 
We shall discuss such choices in section IV. 
\section{Group theoretical determination of mixing}
The physical neutrino mixing matrix $U_{PMNS}\equiv U$ depends on the
structure of $M_\nu$ and $M_l M_l^\dagger$. The latter  can be determined if
the
symmetry $T_l$ as in eq.(1)
is known. We now make an assumption that
$S_\nu$ satisfying
eq.(3) and $T_l$ as in eq.(1) are elements of some discrete subgroup (DSG)
of $SU(3)$ denoted by $G_f$.
The DSG of $SU(3)$ have been classified in
\cite{Miller,Fairbairn:1964sga,Bovier:1980gc}. They
are further studied in
\cite{Luhn:2007yr,Luhn:2007uq,
Escobar:2008vc,Ludl:2009ft,Ludl:2010bj,
Zwicky:2009vt,Parattu:2010cy,Grimus:2010ak,Grimus:2011fk,Grimus:2013apa,
Merle:2011vy}. 
These can be written in terms of
few $3\times 3$ presentation matrices whose multiple products generate
various DSG.
Two main groups series called $C$ and $D$ \cite{Grimus:2013apa}
constitute bulk of the DSG of $SU(3)$. Of these, we shall explicitly study
two infinite groups series $\Delta(3 N^2)$ and $\Delta(6 N^2)$ which are
examples of the type $C$ and $D$ respectively. See
\cite{King:2013vna,Ding:2014ora,Hagedorn:2014wha} for
earlier studies of neutrino mixing using the  groups  $\Delta(3 N^2)$
and $\Delta(6 N^2)$ and neutrino symmetry rather than antisymmetry.

Eq. (1) implies that $T_l$ commutes with $M_lM_l^\dagger$. Thus, the matrix
$U_l$ diagonalizing the former also diagonalizes $M_lM_l^\dagger$ and
corresponds to the mixing matrix among the left handed charged leptons.
Similarly, the matrix $U_\nu$ diagonalizing $M_\nu$ gets related to the
structure of $S_\nu$.
In this way, the knowledge of $S_\nu$ and $T_l$ can be used to determine the
mixing matrix 
\be \label{upmns}
U \equiv U_{\rm PMNS}=U_l^\dagger U_\nu ~.\ee
This is the strategy followed in the general approach and we shall also use
this to determine all possible mixing pattern for a  given $G_f$ consistent
with eqs.(1) and (3).

Not all the groups $G_f$ can admit an $S_\nu$ which will provide a
legitimate
antisymmetry operator $S_\nu$, i.e. an element with eigenvalues specified
by eq.(\ref{dsforms}). Our strategy would be to determine a class of groups
which will have one or more allowed $S_\nu$ and then look for all viable
$T_l$ within these groups. There would be different mixing patterns
associated with each choice of $S_\nu,T_l$ 
and it is possible to determine all of them analytically for $\Delta(3 N^2)$
and $\Delta(6 N^2)$ groups.
\subsection{$\Delta(3 N^2)$}
The $\Delta(3 N^2)$ groups are isomorphic to $Z_N\times Z_N\rtimes Z_3$,
where $\rtimes$ denotes the semi-direct product. The group
theoretical details for $\Delta(3 N^2)$ are discussed in
\cite{Luhn:2007uq,Ishimori:2010au}. For our purpose, it is sufficient to
note
that all the elements of the group are generated from the
multiple product of two
basic generators defined as:
\be \label{cn}
\ba{cc}
F=\left(\ba{ccc}
1&0&0\\
0&\eta&0\\
0&0&\eta^*\\ \ea \right),~~
E=\left(\ba{ccc}
0&1&0\\
0&0&1\\
1&0&0\\ \ea \right)\\ \ea \ee
with $\eta=e^{\frac{2 \pi i}{N}}$. Here $F$ generates one of the $Z_N$
groups and $E$ generates $Z_3$ in the semi-direct product
$Z_N\times Z_N\rtimes Z_3$. The other $Z_N$ group is generated by
$EFE^{-1}$. The above explicit matrices provide a faithful
three dimensional irreducible representation of the group and
multiple products of these matrices therefore generate the entire group
whose elements can be labeled as:
\beqa
\label{elements3nsquare}
 W\equiv W(N,p,q)&=&\left(
\ba{ccc}
\eta^p&0&0\\
0&\eta^q&0\\
0&0&\eta^{-p-q}\\
\ea \right),~
R\equiv R(N,p,q)=\left(
\ba{ccc}
0&0&\eta^p\\
\eta^q&0&0\\
0&\eta^{-p-q}&0\\
\ea \right), \nonumber \\
V\equiv V(N,p,q)&=&\left(
\ba{ccc}
0&\eta^p&0\\
0&0&\eta^q\\
\eta^{-p-q}&0&0\\
\ea \right).\eeqa
All elements of $\Delta(3 N^2)$ are obtained by
varying $p,q$ over the  allowed range $p,q=0,1,2,..,N-1$ in the above
equation. Thus each matrices $W,R,V$ have $N^2$ elements giving in total 
 $3N^2$ elements corresponding to the order of $\Delta(3N^2)$. The
eigenvalue
equation for the $2 N^2$ non-diagonal
elements $R$ and $V$ is simply given by $\lambda^3=1$. These elements
therefore have eigenvalues $(1,\omega,\omega^2)$ with $\omega=e^{\frac{2 \pi
i}{3}}$. These are not in the form of eq.(\ref{dsforms}) required to get
the 
neutrino antisymmetry operator $S_\nu$. Thus $S_\nu$  has to come from the 
$N^2$ diagonal elements. This requires that $N,p,q$ should be such that 
$W(N,p,q)={\rm diag.}(\eta^p,\eta^q,\eta^{-p-q})$ matches the required
eigenvalues $\tilde{S}_\nu$  of $S_\nu$ given by eq.(\ref{dsforms}) or their
permutations. This cannot happen for all the values of variables and one can
easily
identify the viable cases. It is found that
\begin{itemize}
\item $W$ can match any of $\tilde{S}_\nu$ only for even $N$. Thus only
$\Delta(12k^2)$ groups with $k=1,2,....$ contain neutrino antisymmetry
operator $S_\nu$.
\item The eigenvalue set $\tilde{S}_\nu=(1,-1,-1)$ is always contained as a diagonal
generator for all $\Delta(12 k^2)$ groups and can be chosen as
$S_\nu=W(2k,0,k)$. Hence the
texture I with two degenerate and one massless neutrino can follow in any
$\Delta(12 k^2)$. The smallest such group is $\Delta(12)=A_4$ which is one
of the most studied flavour symmetry from other points of
view \cite{Ma:2001dn,Babu:2002dz,Altarelli:2005yx,Gupta:2011ct,Ma:2015pma,
He:2006dk,He:2015gba,Hirsch:2007kh,Dev:2015dha,He:2015afa,He:2015gba}.
\item The set $\tilde{S}_\nu=(\pm i,\pm i,-1)$ arises only for N multiple of
4, i.e. 
in case of groups $\Delta(48 l^2)$, $l=1,2...$. These groups also contain 
a $\tilde{S}_\nu$ satisfying the second of eq.(\ref{dsforms}). Thus textures
$I,II,IV$ are possible for all $\Delta(48 l^2)$ groups.
\item The set $\tilde{S}_\nu=(\lambda,-\lambda^*,-1)$ with $\lambda\not= \pm 1 ,\pm i$
and the associated texture III is viable in $\Delta(12 k^2)$ with $k\geq 3$
\end{itemize}
Let us now turn to the mixing pattern allowed within the $\Delta(12 k^2)$
groups. $S_\nu$ has to be a diagonal operator identified above. Then $T_l$
can be any other diagonal operator $W(2k,p,q)$ or any of $R(2k,p,q)$ or
$V(2k,p,q)$. In the former case, $U_l=\mathbbm{1}$, where $\mathbbm{1}$ denotes a
$3\times 3$ identity matrix. The neutrino
mixing in
this case coincides with $V_\nu$ diagonalizing any of the four textures of
$\tilde{M}_\nu$ giving 
$U_{\rm PMNS}=V_\nu$. None of
the allowed 
$V_\nu$ are suitable to give the correct mixing pattern with a non-zero
$\theta_{13}$. Thus, $T_l$ needs to be any  of the non-diagonal element $R,
V$. The matrices $V_{R,V}$ diagonalizing $R,V$ are given by
\beqa
 \label{vpq}
V_R(N,p,q)&=&{\rm diag.}(1,\eta^q,\eta^{-p})U_\omega~,\nonumber \\
V_V(N,p,q)&=&{\rm diag.}(1,\eta^{-p},\eta^{-p-q})U_\omega^*~,\eeqa
where,
\be
\label{uw}
U_\omega=\frac{1}{\sqrt{3}}\left( \ba{ccc}
1&1&1\\
1&\omega^2&\omega\\
1&\omega&\omega^2\\
\ea \right)~.\ee
The final mixing matrix depends upon the choice of specific texture for
$\tilde{M}_\nu$. Consider the texture I which arises within all the
$\Delta(12 k^2)$ groups. $U_\nu=V_\nu$ in this case is given by
eq.(\ref{unu1}) and $U_{\rm PMNS}=V_{R,V}^\dagger V_\nu$.
Since a neutrino pair is degenerate, the solar mixing
angle $\theta_{12}$ remains undetermined in the symmetry limit. 
This is reflected by
the presence of an unknown angle $\phi$ in eq.(\ref{unu1}). 
In this case, the neutrino mass hierarchy is inverted and the  third column
of
$U_{\rm PMNS}\equiv U$ needs to be identified with 
the massless state. It 
is independent of the angle $\phi$. We get for $T_l=R(N,p,q)$,
\be\label{ui3}
U_{i3}=\frac{1}{\sqrt{3}}\left(c e^{-i\beta}\eta^{p+q}-s ,c
\omega e^{-i\beta}\eta^{q+p}-s\omega^2 ,c \omega^2
e^{-i\beta}\eta^{q+p}-s \omega
\right)^T~ \ee
with $\eta=e^{\frac{\pi i}{k}}$ for the group $\Delta(12 k^2)$.
$p,q$ take
discrete values $0....2k-1$ in above equation while $\beta$ and $\theta$
are unknown quantities appearing in the neutrino mixing matrix 
eq.(\ref{unu1}). The entries in $U_{i3}$ can be
permuted by reordering the eigenvalues of $T_l$. We will identify the
minimum of $|U_{i3}|^2$ with $s_{13}^2$. If the   minimum of the
remaining two is identified 
with $c_{13}^2 s_{23}^2$ then one will get a solution with the atmospheric
mixing angle $\theta_{23}\leq 45^\circ$. In the converse case, one will
get a solution $\geq 45^\circ$. The experimental
values of the leptonic angles are determined through
fits to neutrino oscillation data
\cite{Capozzi:2013csa,Forero:2014bxa,Gonzalez-Garcia:2014bfa}.
Throughout, we shall specifically use the fits presented in
\cite{Capozzi:2013csa} for definiteness. The texture I corresponds to
the inverted hierarchy and the best
fit values and 3$\sigma$ ranges 
appropriate for this case are given \cite{Capozzi:2013csa} by:
\beqa\label{fits}
\sin^2\theta_{12}&=&0.308~(0.259-0.359)~,\nonumber
\\
\sin^2\theta_{23}&=&0.455~
(0.380-0.641)~,\nonumber \\
\sin^2\theta_{13}&=&0.0240~(0.0178-0.0298)~.\eeqa
Let us
mention salient features of results following from  eq.(\ref{ui3})
\begin{itemize}
\item It is always possible to obtain correct $\theta_{13},\theta_{23}$ by
choosing unknown quantities $\theta$ and $\beta$ of $\tilde{M}_\nu$.
This
should be contrasted with situation found in \cite{Joshipura:2014qaa} which
used neutrino
symmetry instead of antisymmetry to obtain a degenerate pair of neutrinos.
As
discussed there, none of the $\Delta(3N^2)$ groups could 
simultaneously account for the  values of
$\theta_{13},\theta_{23}$ within 3$\sigma$.
\item It is possible to obtain more definite predictions by choosing 
specific values of $\theta$ and or $\beta$. In contrast to $\theta$ and
$\beta$ which are unknown, the choice of $p,q$ is  dictated by the choice of
$T_l$ and it is possible to consider any specific choice of $p,q$ in the
range $0,...N-1$. Consider a very specific choice of real
$\tilde{M}_\nu$
,i.e. $\beta=0$ and a residual symmetry $T_l=E^2$ corresponding to putting 
$p=q=0$ in eq.(\ref{ui3}). This equation in
this case gives a prediction $|U_{23}|=|U_{33}|$ which holds for
all values of $\theta$. 
This relation is equivalent to a maximal 
$\theta_{23}$ which lies within the 1$\sigma$ range of 
the global fits \cite{Capozzi:2013csa}. $\theta$ then can be chosen to get
the correct
$\theta_{13}$. Since the specific choice $p=q=0$ is allowed within all
the $\Delta(12 k^2)$ groups, all of them  can predict the maximal
$\theta_{23}$ and can accommodate correct $\theta_{13}$.
\item   The relation $|U_{23}|=|U_{33}|$ does not hold for a
complex $\eta^{p+q}$ even if $\beta=0$. Such choices of $T_l$ give
departures from maximality in $\theta_{23}$. It is then possible to
reproduce 
both the angles correctly by choosing $\theta$.
This is non-trivial since a single unknown $\theta$ determines both
$\theta_{13}$ and $\theta_{23}$ for a specific choice of group (i.e. $N$)
and a residual symmetry
$T_l$ (i.e. $p$ and $q$). The resulting prediction can be worked out
numerically 
by varying $p,q,N$ over the allowed integer values and $\theta$ over 
continuous range from $0$ to $2 \pi$. Values of $s_{23}^2$ obtained this way
are  depicted in Fig.(1). This is obtained by requiring that $s_{13}^2$
lies within the allowed 1$\sigma$ range. The phase $\beta$ is put to zero.
It is seen form the Figure that 
all the $\Delta(12 k^2)$ groups always allow maximal $\theta_{23}$ as
already discussed. But solutions away from maximal are also possible for
$k\geq 4$. The minimal group capable of
doing this is $\Delta(192)$. The next group $\Delta(300)$ can lead to near
to the 
best fit values of the parameters. Specifically the choice 
$T_l=R(10,0,7)$, $S_\nu=W(10,0,5)$ within the group and $\theta\sim 54.3^\circ$ gives 
$s_{13}^2\sim 0.024$ and
$s_{23}^2\sim 0.442$ to be compared with the best fit values 
$0.024$ and $0.455$ in \cite{Capozzi:2013csa}.
\item  $p,q$ can only be  zero or 1 and $\eta$ is real for the smallest
group $\Delta(12)=A_4$. In this case, one 
immediately gets the prediction $\theta_{23}=\frac{\pi}{4}$ for
$\beta=0$.  $\mu$-$\tau$ symmetry is often used to predict 
the maximal $\theta_{23}$. This is not even contained in $A_4$ which has
only
even permutations of four objects. Still the use of
antisymmetry
rather than symmetry allows one to get the  maximal $\theta_{23}$ and it
also
accommodates a non-zero $\theta_{13}$ within $A_4$.  This should be
contrasted with the situation obtained in case
of the use of symmetry condition eq.(2) instead of (3). It is known that
in this case $A_4$ group gives democratic value $\frac{1}{3}$ for
$s_{13}^2$, see for example \cite{deAdelhartToorop:2011re}.
\end{itemize}
\begin{figure}[!ht]
\centering
\includegraphics[width=0.5\textwidth]{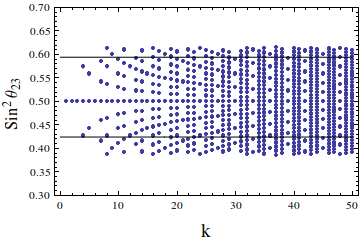}
\caption{Predictions for $\sin^2\theta_{23}$ for the groups $\Delta(12
k^2)$ as a function of $k$ when $\sin^2\theta_{13}$ is allowed to vary
within the 1$\sigma$ range as obtained through global
fits in \cite{Capozzi:2013csa}. Horizontal lines show 1$\sigma$
limits on $\sin^2\theta_{23}$.}
\label{fig1}
\end{figure}

We now argue that the other three textures though possible within $\Delta(12
k^2)$ groups do not give the  the correct mixing pattern. Texture II has one
massless and in general two non-degenerate neutrinos. This texture can
give both the  normal  and the inverted hierarchy.
The mixing matrix $V_\nu$ is block-diagonal with
a $2\times 2$ matrix giving mixing among two massive states. Given this
form for $V_\nu$ and a general $U_l$ as given in eq.(\ref{vpq}), one finds
that the case with inverted hierarchy leads to the prediction
$\sin^2\theta_{13}=\frac{1}{3}$ while the normal hierarchy gives instead 
$\cos^2\theta_{13}\cos^2\theta_{12}=\frac{1}{3}$. Neither of them come
close to their experimental values. 

The texture (III) having  degenerate pair corresponds to the inverted
hierarchy. $V_\nu$ in this case is block diagonal with an unknown solar
angle. Given the most general form, eq.(\ref{vpq}) for $V_l$ one 
obtains once again the wrong prediction $\sin^2\theta_{13}=\frac{1}{3}$
ruling out this texture as well. Likewise, texture IV also gets
ruled out. This corresponds to a diagonal $\tilde{M}_\nu$ with 
$V_\nu=\mathbbm{1}$ and $|U_{\rm PMNS}|=|U_l|$ has the universal structure
$|U|=\frac{1}{3}\mathbbm{1}$.

To sum up, all the groups $\Delta(12k^2)$ contain a neutrino antisymmetry
operator $S_{\nu}$ and allow a neutrino mass spectrum with two degenerate
and one massless neutrino and can reproduce correctly two of the mixing
angles $\theta_{13},\theta_{23}$. The  values for the solar angle and the
solar scale have to be generated by small perturbations within these group.
We shall study an example based on the minimal group $A_4=\Delta(12)$ in
this category in section V.
\subsection{$\Delta(6N^2)$ groups}
$\Delta(6 N^2)$ groups are isomorphic to $Z_N\times Z_N\rtimes
S_3$  with $N=1,2,3...$. The $S_3$ group in the semi-direct product is
generated by $E$ in eq.(\ref{cn}) and a matrix 
\be\label{mt}
G=-\left(\ba{ccc}
1&0&0\\
0&0&1\\
0&1&0\\
\ea \right)~.\ee
The matrices $E,F,G$ provide a faithful irreducible representation
of $\Delta(6N^2)$ \cite{Escobar:2008vc} and generate the entire group
with $6N^2$ elements. $3 N^2$ elements generated by $E,F$ give the
$\Delta(3 N^2)$ subgroup. The 
additional $3N^2$ elements are generated from the multiple products of $G$
with elements of $\Delta(3 N^2)$.
These new elements can be parameterized by:
\beqa
\label{elements6nsquare}
S\equiv S(N,m,n)=-\left(
\ba{ccc}
\eta^m&0&0\\
0&0&\eta^n\\
0&\eta^{-m-n}&0\\
\ea \right)~~,&~~
T\equiv T(n,m,n)=-\left(
\ba{ccc}
0&0&\eta^n\\
0&\eta^m&0\\
\eta^{-m-n}&0&0\\
\ea \right)~,\nonumber\\
~~~~~~~~~~~~~~~
U\equiv U(n,m,n)=-\left(
\ba{ccc}
0&\eta^n&0\\
\eta^{-m-n}&0&0\\
0&0&\eta^m\\
\ea \right) ~\!\!\!\!\!~.&\eeqa
Here  $0\leq (m,n)<N-1$.
Since $\Delta(3 N^2)$ is a subgroup of $\Delta(6 N^2)$, the neutrino mass
and mixing patterns derived in the earlier section 
can also be obtained here. But the new elements $S,T,U$ allow more
possibilities now. In particular, they allow 
more elements which can be used as neutrino antisymmetry $S_\nu$. To see
this, note that the eigenvalues of
$S,T,U$ are given by $(\eta^{-m/2},-\eta^{-m/2},-\eta^m)$. This can have
the required form, eq.(\ref{dsforms}) when
$m=0$ or $m=N/2$. The eigenvalues in respective cases are 
$(1,-1,-1)$ or $(-i,i,1)$
and one gets the textures I or IV by using 
any of  $S,T,U$ as neutrino antisymmetry with $m=0$ and $m=N/2$
respectively.
Similarly, possible choices of the charged lepton
symmetry $T_l$ also increases. It can be any of the six types of elements:
$W$,$R,V$ as before or $S,T,U$. Important difference compared to 
$\Delta(3 N^2)$ is that the texture I can now be obtained  for both odd and
even values of $N$ by choosing any  of the $S,T,U$ with $m=0$ as neutrino
antisymmetry.
Texture IV still requires $m=N/2$ and hence even $N$ for its
realization. We determine mixing matrix $U$ for each of these 
textures and discuss them in turn.
\subsubsection{Texture I}
The residual anti symmetries which lead to texture I  can be either 
(1) $S_\nu=W(2k,0,k)$ or (2) $S_\nu=P(N,0,n)$ where
$P=S,T,U$.
The residual symmetry $T_l$ 
of $M_lM_l^\dagger$ can be any elements in the group which we divide in
three classes:
$(A)~ T_l=
W(N,p,q)$, $(B)~ T_l=P(N,p,q)$ and $(C)~
T_l=Q(N,p,q)$. Here and in the following, we  use symbols $P$
and $Q$ to collectively denote $P=S,T,U$ and $Q=R,V$. We use the basis as specified in eqs.(\ref{elements6nsquare},\ref{elements3nsquare}) for $S_\nu,T_l$. Then the
neutrino mixing matrix is given by $U_\nu=V_\nu$ in case (1) while it 
is given by  $U_\nu=V_P(N,0,n)V_\nu$ in case (2). This
follows by noting that the texture $\tilde{M}_\nu$ given in eq.(\ref{text1})
holds in a basis with diagonal $S_\nu$ but $S_\nu$ in the chosen group
basis of eq.(\ref{elements6nsquare}) is non-diagonal  in case (2).
The neutrino mass matrix in this basis is thus given by
$M_\nu=V_P^*\tilde{M}_\nu V_P^\dagger $  where 
$V_P$ diagonalizes $P(N,0,n)$. The matrix $U_\nu$ which
diagonalizes $M_\nu$
is then given by $U_\nu=V_P(N,0,n)V_\nu$ where  $V_\nu$ diagonalizes
$\tilde{M}_\nu$. Explicitly,
$V_P^\dagger(N,p,q) P(N,p,q) V_P(N,p,q)={\rm diag.}
(\eta^{-p/2},-\eta^{-p/2},-\eta^{p})$ with\\
\beqa \label{vstu}
V_S(N,p,q)=\frac{1}{\sqrt{2}}\left( 
\ba{ccc}
0&0&\sqrt{2}\\
1&\eta^{q+p/2}&0\\
-\eta^{-q-p/2}&1&0\\
\ea \right)&~;&
V_U(N,p,q)=\frac{1}{\sqrt{2}}\left(
\ba{ccc}
1&\eta^{q+p/2}&0\\
-\eta^{-q-p/2}&1&0\\
0&0&\sqrt{2}\\
\ea \right)~,\nonumber \\
V_T(N,p,q)=\frac{1}{\sqrt{2}}\left(
\ba{ccc}
1&\eta^{q+p/2}&0\\
0&0&\sqrt{2}\\
-\eta^{-q-p/2}&1&0\\
\ea \right)&.& \eeqa
We have chosen the ordering of columns of $V_P$ in such a way that the first
column always corresponds to the eigenvalue
$\eta^{-p/2}$. With this ordering one gets the texture I given in
eq.(\ref{text1}) when $P(N,0,n)$ is used as neutrino antisymmetry. 

The matrices $U_l$ diagonalizing $T_l$ in three cases above are given in the
same basis by $U_l=\mathbbm{1},V_P (N,p,q),V_Q(N,p,q)$ in cases
(A),(B),(C) respectively where $V_Q$ are given in eq.(\ref{vpq}). Thus we have six (four)  different choices for
$U_l$ ($U_\nu$) giving  in all 24 leptonic mixing matrices $U_{\rm PMNS}$.
We list these choices and the corresponding $U_{\rm PMNS}$ matrices in Table
I.
\begin{table}[!ht]
\begin{small}
\begin{center}
\begin{tabular}{|c|c|c|c|c|c|}
\hline
\hline
Case &$S_\nu$&$T_l$&$U_l$ &$U_\nu$&$U_{PMNS}$\\
\hline
1A&$W(2k,0,k)$&$W(N,p,q)$&$\mathbbm{1}$&$V_\nu$&$V_\nu$\\
\hline
1B&$W(2k,0,k)$&$P(N,p,q)$&$V_P(N,p,q)$&$V_\nu$&$V_P^\dagger(N,p,q)V_\nu$\\
\hline
1C&$W(2k,0,k)$&$Q(N,p,q)$&$V_{Q}(N,p,q)$&$V_\nu$&
$V_{Q} ^\dagger (N,p,q) V_\nu$\\
\hline
2A&$P(N,0,n)$&$W(N,p,q)$&$\mathbbm{1}$&$V_P(N,0,n)V_\nu$&$V_P(N,0,n)V_\nu$\\
\hline
2B&$P(N,0,n)$&$P^\prime(N,p,q)$&$V_{P^\prime}(N,p,q)$&$V_P(N,0,
n)V_\nu$&$V_{P^\prime}^\dagger(N , p ,
q)V_P(N , 0 , n)V_\nu$\\
\hline
2C&$P(N,0,n)$&$Q(N,p,q)$&$V_{Q}(N,p,q)$&$V_P(N,0,
n)V_\nu$&$V_ {Q}^\dagger(N,p,q)V_P(N,0,n)V_\nu$\\
\hline
\hline
\end{tabular}
\end{center}
\end{small}
\caption{All possible choices of the residual symmetries $S_\nu$ and $T_l$
within
$\Delta(6N^2)$ groups and the corresponding PMNS mixing matrices.
$P,P^\prime$ collectively denote any of $S,T,U$ defined in the text. $Q$
denotes $R$ and $V$. The mixing matrices $V_P,V_Q$ and $V_\nu$ appearing
above are given in eq.(\ref{vstu}), eq.(\ref{vpq})  and eq.(\ref{unu1})
respectively.}
\end{table}

Not all of  24 mixing matrices listed in Table I give
independent predictions for the third column of $U_{PMNS}$ which determines
$s_{13}$ and $s_{23}$. We discuss the independent ones below.

The choice (1A) giving $U_{PMNS}=V_\nu$ has one of the entries zero and thus
cannot lead to correct $\theta_{13}$ or $\theta_{23}$. 
The choice (1C) involves only elements belonging to the $\Delta(3N^2)$
subgroup and its predictions are already discussed in the previous
section. The remaining choices give new predictions.\\
\noindent The case (1B) leads to three different $U_{PMNS}$. One obtained
with $T_l=S(N,p,q)$  contain a zero entry in the third column and can be
used only as a zeroeth order choice. One gets the following result in (1B)
if $T_l=T(N,p,q)$
\be
\label{1B}
|U_{23}|^2=\frac{c^2}{2}~,~|U_{33}|^2=\frac{c^2}{2}~,~|U_{13}|^2=s^2
~.\ee
The ordering of the entries $|U_{i3}|^2$ can be changed by rearranging
the eigenvectors of  of $T_l$ appearing in $U_l$. We have chosen here and
below  an ordering  which is consistent with the values of the parameters
$s_{13}^2,s_{23}^2$ when $U$ is equated with the standard form of the
mixing matrix.The result in the third case with $T_l=U(N,p,q)$ can be
obtained
from above by the replacement $s\leftrightarrow c$. All the three entries
above follow for all the choices of $p,q$ and the phase $\beta$. The case
(1B) in this way gives a  universal prediction. Two of the
$|U_{i3}|^2$ are equal within this choice and they correspond to
$c_{13}^2c_{23}^2$ and $c_{13}^2 s_{23}^2$.
Equality of the two  then 
implies a $\theta$ independent prediction $\theta_{23}=\frac{\pi}{4}$.
$s_{13}^2$ in
the above case is then given by $s^2$ and can match the experimental
value with
appropriate choice of the unknown $\theta$. Since the choice of $S_\nu$
within
(1B) is possible only for even $N$ it follows that all the groups
$\Delta(24k^2)$ lead to a prediction of the maximal atmospheric mixing angle
and can accommodate the correct $\theta_{13}$.

The choice (2A) also gives the same result for $|U_{i3}|^2$  as (1B) with an
important
difference. The neutrino residual symmetry used in this choice is allowed 
for all $N$ and not necessarily $N=2k$. Thus one gets a universal prediction
of the maximal $\theta_{23}$ for all $p,q,\theta,\beta$ within
all $\Delta(6N^2)$ groups.The smallest group in
this category is the permutation group $S_4=\Delta(24)$ which contain 
symmetries appropriate for both the cases $(1B)$ and $(2A)$. 
      
There are two independent structures within nine possible choices contained
in case (2B). The example of the  first one is provided by the choice  
$S_\nu=S(N,0,n)$ and $T_l=S(N,p,q)$. The elements in the third column of 
mixing matrix are given in this case by
\beqa \label{2Bss}
|U_{23}|^2&=&\frac{1}{4}s^2|\eta^n-\eta^{q+p/2}|^2~,\nonumber \\
|U_{33}|^2&=&\frac{1}{4}s^2|\eta^{-n}+\eta^{-q-p/2 } |^2~,\nonumber\\
|U_{13}|^2&=&c^2
~.\eeqa
While this choice does not give universal prediction as in the case (1B) discussed above it still leads to 
a prediction for $\theta_{23}$ which is independent of the unknown angle $\theta$ and phase $\beta$:
$$\tan^2\theta_{23}~~{\rm or}~~
\cot^2\theta_{23}=\frac{|\eta^n-\eta^{q+p/2}|^2}{|\eta^{-n}+\eta^{-q-p/2}|^2
}$$
This follows from eq.(\ref{2Bss}) when $|U_{13}|^2$ is identified with
$s_{13}^2$. The predicted $\theta_{23}$ now
depends only  on the  group theoretical factors $N,p,q,n$.

Unlike (1B), both the maximal and non-maximal values are allowed for
$\theta_{23}$ in this case. The former occurs whenever
$\cos \frac{2\pi(n-q-p/2)}{N}=0$. The latter occurs for
other choices. It
is possible to find values of parameters  which lead to a non-maximal $\theta_{23}$  within
the experimental limits. 
The minimal such choice occurs 
for $N=7$, i.e. the group $\Delta( 294)$ which leads as shown in Table II
to a $\sin^2\theta_{23}$
within the 2$\sigma$ range as given in \cite{Capozzi:2013csa}. The
next example of
the group $\Delta(486)$ fairs slightly better.

The other prediction of the case (2B) is obtained with
$S_\nu=S(N,0,n)$ and $T_l=T(N,p,q)$. One obtains in this case
\beqa \label{2Ast}
|U_{23}|^2&=&
\frac{1}{4}|\sqrt{2}c e^{-i \beta}+s \eta^{q+p/2}|^2~,\nonumber\\
|U_{33}|^2&=&
\frac{1}{4}|\sqrt{2}c e^{-i \beta}-s \eta^{-q-p/2}|^2~,\nonumber\\
|U_{13}|^2&=&\frac{1}{2}s^2~.\eeqa
In this case, $\theta_{23}$ is necessary non-maximal if $\theta_{13}$ is to be small but non-zero.
We may identify,
$|U_{13}|^2$ with $s_{13}^2$ and fix  $s^2=2 s_{13}^2$. This determines the
other two entries of $|U_{i3}|^2$ for a given $p,q,\beta$. For $p=q=\beta=0$
one obtains 
$\sin^2\theta_{23}$ either $0.345$ or $0.655$. Thus
all the $\Delta(6N^2)$ groups with this specific choice give results 
close to the 3$\sigma$ range in the global fits.
This prediction can be improved by turning on $\beta$ or choosing different
$T_l$. An example based on the group
$\Delta(150)$ giving $\sin^2\theta_{23}$ close to  the best fit value
\cite{Capozzi:2013csa}  is shown in the table.
%%%%%%%%%%%%%%%%%%%%%%%%%%%%%%%%%%%%%%%%%%%%%%%%%%%%%%%%%%%%%%%%%%%%%%%%%%%%
%%%%%%%%%%%%%%%
\begin{table}[!ht]
\begin{small}
\begin{center}
\begin{tabular}[c]{c|c|c|c}
\hline
\hline
Group &$T_l$  &$S_\nu$ 
&Predictions\\
\hline
\hline
$\Delta(24 k^2)$&~~ $T(2k,p,q)$~~& ~~$W(2k,0,k)$&Maximal
$\theta_{23}$ for all $\beta,p,q,n$\\
&~& ~~&$\sin^2\theta_{13}=\sin^2\theta$\\
\hline
$\Delta(6N^2)$&~~ $W(N,p,q)$~~& ~~$P(N,0,n)$&Maximal
$\theta_{23}$ for all $\beta,p,q,n$\\
&~& ~~&$\sin^2\theta_{13}=\cos^2\theta$\\
\hline
$\Delta(6N^2)$&~~ $S(N,p,q)$~~& ~~$S(N,0,n)$&Maximal
$\theta_{23}$ for  all  $\beta$, $\frac{|n-q-p/2||}{N}=\frac{(2l+1)}{4}$\\
&~& ~~&$\sin^2\theta_{13}=\cos^2\theta$\\
\hline
$\Delta(294)$&~~ $S(7,0,2)$~~& ~~$S(7,0,0)$&$s_{23}^2=0.39$ or $0.61$ for
all $\theta,\beta$\\
&~& ~~&$\cos^2\theta=s_{13}^2$\\
\hline
$\Delta(486)$&~~ $S(9,0,2)$~~& ~~$S(9,0,0)$&$s_{23}^2=0.41$ or $0.59$ for
all $\theta,\beta$\\
&~& ~~&$\cos^2\theta=s_{13}^2$\\
\hline
$\Delta(6N^2)$&~~ $S(N,0,0)$~~& ~~$T(N,0,n)$&$\sin^2\theta=2 s_{13}^2$\\
&~& ~~&$s_{23}^2=0.345$ for $\beta=0$ and the best fit $s_{13}^2$\\
\hline
$\Delta(150)$&~~ $S(5,2,3)$~~& ~~$T(5,0,0)$&$\sin^2\theta=2 s_{13}^2$\\
&~& ~~&$s_{23}^2=0.452$ for the  best fit $\theta_{13}$\\
\hline
$\Delta(6N^2)$&~~ $R(N,0,0)$~~& ~~$S(N,0,0)$&$s_{23}^2=1/2$ for $\beta=0$
and  all
$\theta$\\
&~& ~~&$s_{13}^2=0.028$ for $\beta=0$, maximal $\theta$\\
\hline
$\Delta(150)$&~~ $R(5,3,1)$~~& ~~$S(5,0,0)$&$s_{23}^2=0.484$,
$s_{13}^2=0.022 $\\
&~& ~~&for $\beta=0$ and $\theta\sim 80^\circ$\\
\hline
\hline
\end{tabular}
\caption{Some illustrative predictions of the mixing angles
$\sin^2\theta_{13}$ and $\sin^2\theta_{23}$ using $\Delta(6N^2)$ groups as
flavour symmetry. $\sin^2\theta_{12}$ remains undetermined due to
degeneracy in two of the masses in all these cases.}
\end{center}
\end{small}
\end{table}
%%%%%%%%%%%%%%%%%%%%%%%%%%%%%%%%%%%%%%%%%%%%%%%%%%%%%%%%%%%%%%%%%%%%%%%%%%%%
%%%%%%%%%%%%%%

Predictions of the case (2C) can also be similarly worked out. Six different
$U_{\rm PMNS}$ are associated with this choice but not all give different predictions for the third column.
One of the independent structures corresponds to choosing $T_l=R(N,p,q)$ and
$S_\nu=S(N,0,n)$. The $U_{\rm PMNS}=V_R^\dagger V_S V_\nu$ gives
\beqa \label{predict2C}
|U_{13}|^2&=&\frac{1}{6}\left|
-s(\eta^p+\eta^{n-q})+\sqrt{2} ce^{-i\beta}\right|^2 ,\nonumber \\
|U_{23}|^2&=&\frac{1}{6}\left|
-s(\eta^p\omega+ \eta^{n-q}\omega^2)+\sqrt{2} c e^{-i\beta}\right|^2
,\nonumber \\
|U_{33}|^2&=&\frac{1}{6}\left|
-s(\eta^p \omega^2+\eta^{n-q} \omega)+\sqrt{2} c e^{-i\beta}\right|^2
~.\eeqa
Rest of the choices within (2C) differ from the above only in the powers of $\eta$. Their predictions 
can be obtained from the above by choosing different values of $p,q,n$. 

Eq.(\ref{predict2C})  gives $s_{13},s_{23}$ withing
3$\sigma$ range for a
suitable choice of $N,p,q,\theta,\beta$.
In particular, one predicts a maximal $\theta_{23}$ if $p=q=\beta=0$ as
in the earlier cases. But now the
maximal value of $\theta$ also becomes a viable choice for all
the groups $\Delta(6N^2)$. This makes the choice in this class particularly interesting since such value of $\theta$ can be 
forced by some additional symmetry. With the choice $T_l=E^2$ corresponding to $p=q=0$,
eq.(\ref{predict2C}) gives for $n=\beta=0$,
$$\theta_{23}=\frac{\pi}{4}~,~ s_{13}^2=\frac{1}{3}|c-\sqrt{2}s|^2~.$$
The maximal value of $\theta$ then leads to $s_{13}^2\sim 0.029$ which is close to
2$\sigma$ range as obtained in \cite{Capozzi:2013csa}. One can obtain
a better solution with a different choice  for $p$ and $q$ and $\theta$. One
particular solution based on the group $\Delta(150)$ is shown in the table.
\subsubsection{Texture IV} 
The diagonal texture IV given in eq.(\ref{text4}) can be realized in
$\Delta(6 N^2)$ for even $N$ with the choice $S_\nu=P(2k,k,n)$.
This texture has two non-degenerate and one massless neutrino. Thus both the
normal and the inverted hierarchies are possible. The massless state has to
be identified with the third (first)
column of the mixing matrix for the inverted (normal) hierarchy. The
neutrino mixing matrix $U_\nu$ in this case is given by the matrix which
diagonalizes $P(2k,k,n)$. This is given for the inverted hierarchy by
$V_{P}(2k,k,n)$ as defined in eq.(\ref{vstu}). For the normal hierarchy, one
instead gets $U_\nu=V_{P}(2k,k,n) Z_{13}$ where $Z_{13}$ exchanges the
first and the third column of of the mixing matrix obtained in case of the 
inverted hierarchy. Possible choice of $T_l$ can be any of the six types of
generators and corresponding mixing matrices $U_l$ are the same as given in
Table I with the choice (2A),(2B),(2C). It is then straightforward to work
out the final mixing matrices $U_{PMNS}$. As the massless state in the
basis with diagonal $S_\nu$ is given by $(1,0,0)^T$ and its cyclic
permutation
for $S_\nu=S,T,U$, the third column of the $U_{PMNS}$ is given by
$(U_{PMNS})_{i3}=(U_l^\dagger)_{i1},(U_l^\dagger)_{i2},(U_l^\dagger)_{i3}$
when $S_\nu=S,T,U$. It follows from the structure of $U_l$ that the third
column has either one or two zero entries or all elements have equal
magnitudes. The same applies to the first column of $U_{PMNS}$ in case of
the normal hierarchy. In either case, the texture IV cannot give
phenomenologically consistent result at the zeroeth order.
\section{More predictive scenario: $Z_2\times Z_2$ symmetry}
We have assumed so far that the flavour antisymmetry $S_\nu$ is the only
invariance 
of the neutrino mass matrix. This fails in determining
$\tilde{M}_\nu$
completely in case of the texture I which still has two unknown quantities
$\theta$ and
$\beta$ . We give here example of an enlarged residual symmetry of 
$M_\nu$ which serves to determine $\tilde{M}_\nu$ completely apart
from 
an overall complex mass scale. We use an additional symmetry $S_\nu^\prime$
commuting with $S_\nu$ for this purpose. It should be such that
$S_\nu,S_\nu^\prime$ and
$T_l$
together are contained in some $G_f$.  $M_\nu$ may be
antisymmetric with respect to transformation by $S_\nu^\prime$ also. In this
case,
it will be symmetric with respect to
the product $S_\nu S_\nu^\prime$. Instead we assume that $S_\nu^\prime$ is a
symmetry of  $M_\nu$, i.e.
\be
\label{snuprime1}
S_\nu^{\prime T} M_\nu S_\nu^\prime=M_\nu~. \ee
We can transform above equation to the basis with a diagonal $S_\nu$ by
defining $\tilde{S_\nu^\prime}\equiv V_{S_\nu}^\dagger S_\nu^\prime
V_{S_\nu}$.
In
this
basis, we get
\be
\label{snuprime2}
\tilde{S_\nu}^{\prime T}{\tilde{M}_\nu}\tilde{
S_\nu}^\prime=\tilde{M}_\nu~,.
\ee
As before, we demand $\tilde{S_\nu^\prime}$ to be contained in
$SU(3)$.
If
it is
diagonal, then $\tilde{S_\nu^\prime}={\rm
diag.}(\lambda_1,\lambda_2,\lambda_1^*\lambda_2^*)$ with $\lambda_{1,2}$
being roots of unity. Then eq.(\ref{snuprime2}) when applied to
$\tilde{M}_\nu$
in
eq.(\ref{unu1}) implies that either $\tilde{S_{\nu}}^\prime$ is
proportional to identity or $s=0$ or  $c=0$. A
non-trivial
prediction can be obtained if $\tilde{S_\nu}'$ is non-diagonal. Since
$\tilde{S_\nu}={\rm diag.} (1,-1-1)$, a general $\tilde{S_\nu}^\prime$
commuting
with
$\tilde{S_\nu}$
should have a block diagonal structure with the lower $2\times 2$ block
non-trivial. This block gets further restricted from the requirement that 
$S_\nu,S_\nu^\prime,T_l$ are elements of some discrete group $G_f$. These
requirements can be met within the already considered groups $\Delta(6
N^2)$.

Consider the group $\Delta(12 k^2)$. The choice
$S_\nu=\tilde{S_\nu}=W(2k,0,k)={\rm diag.}( 1,-1,-1)$
within it leads to texture I as already discussed. This commutes with all
the discrete symmetries having a general form $S(M,m,n)$ as in
eq.(\ref{elements6nsquare}).
Thus a viable choice for  $S_\nu^\prime$ is provided by
$S_\nu^\prime=S(M,m,n)$.
Note that 
since $S_\nu$ is already diagonal,
$\tilde{S_\nu}^\prime=S_\nu^\prime=S(M,m,n)$.
Then eq.(\ref{snuprime2}) and the form of $\tilde{M}_\nu$ implies a
restriction:
\be \label{theta}
m=0~~~,~~~\theta=\pm\frac{\pi}{4}~~~~,~~~ \beta=\frac{2  \pi n}{M}~\ee
which fixes the unknown angle $\theta$ and phase $\beta$. Mixing pattern can
be
determined by choosing appropriate $T_l$ and let us choose
$T_l=R(N,p,q)$. Since both $T_l$ and $S_\nu$ are contained in $\Delta(12
k^2)$ mixing pattern is determined by the corresponding  eq.
(\ref{ui3}) but now with
$\theta$ and $\beta$ satisfying eq.(\ref{theta}) which follows from the
inclusion of $S_\nu^\prime$ as a residual symmetry. We can vary $p,q,M,N,n$ in eq(\ref{ui3}) and look for
a viable choice. Consider $M=N$ in which case $T_l,S_\nu,S_\nu^\prime$ are contained in
$\Delta(6 N^2)$. By varying $p,q,N$ one finds that the 
minimum group  giving acceptable $\theta_{13},\theta_{23}$ is $\Delta(600)$ corresponding to
$N=10$.
One possible set of residual symmetries  within $\Delta(600)$  is given by
$$S_\nu=W(10,0,5)~~,~T_l=R(10,4,0)~~,~S_\nu^\prime=S(10,0,0)~.$$
With this choice, the $S_{\nu}^\prime$ coincides with the $\mu$-$\tau$
symmetry and  eqs.(\ref{ui3},\ref{theta}) give a prediction
$$ s_{13}^2\approx 0.029~~~~~~,~~~~~~~s_{23}^2\approx 0.38~~{\rm
or}~~0.62$$
to be compared with the 3$\sigma$ region given in
eq.(\ref{fits}).
\section{An $A_4$ model with flavour antisymmetry}
Our discussion so far has been at the group theoretical level. We now
present an explicit realization of flavour antisymmetric neutrino
mass matrix using $A_4$ as an example. $A_4$ has been
extensively used for several different purposes, for obtaining degenerate
neutrinos
\cite{Ma:2001dn,Babu:2002dz}, to realize tri-bimaximal mixing
\cite{He:2006dk,Altarelli:2005yx} for obtaining maximal CP phase
$\delta$
\cite{Gupta:2011ct,Ma:2015pma,Dev:2015dha,He:2015gba,He:2015afa}
or to obtain
texture zeros \cite{Hirsch:2007kh} in the leptonic  mass matrices. As we
discuss
here, it also provides a viable alternative to get a massless and two quasi
degenerate neutrinos with correct mixing pattern. In the following, we
discuss the required symmetry, Higgs content and identify the vacuum needed
to obtain antisymmetry. We also discuss possible perturbations which can
split the degenerate pair and lead to the solar scale and mixing angle.
The aim is not to construct a detailed model but to illustrate how the
basic proposal of the paper can be used for construction of a model.

The group theory of $A_4$ is discussed extensively in many papers. We
shall not elaborate on it. We follow the basis choice as given for example
in \cite{He:2006dk}. In this basis, all the  12 elements of $A_4=\Delta(12)$
can be generated
from the two elements  $E$ and  $F$ defined in eq.(\ref{cn}) with
$\eta=-1$.

We consider supersymmetric model with MSSM extended by a triplet Higgs
field $\Delta$. The standard doublets $H_u,H_d$ and $\Delta$ are $A_4$
singlets. We
use two flavon fields $\chi_\nu$ and $\chi_e$ to break $A_4$ and generate
the flavour structures. Both these fields as well as the three generations of the leptonic doublets
$l_L$ transform as triplets of $A_4$. The right handed charged leptons
transform as  $(e_R,\mu_R,\tau_R)\sim(1,1^\prime,1^{\prime\prime})$
representation of $A_4$. We also
impose an additional $Z_3$ symmetry under which $(l_L,\chi_\nu)\rightarrow
\omega(l_L,\chi_\nu)$ and
$\chi_e\rightarrow \omega^2 \chi_e$. All other fields are assumed singlet
under the $Z_3$. The charged lepton masses arise from the following
superpotential
\be \label{wl}
W_l=\frac{H_d}{M}\left(~ h_e (l_L
\chi_e)_1e_R+h_\mu(l_L
\chi_e)_{1^{\prime\prime}}\mu_R+h_\tau (l_L
\chi_e)_{1^\prime} \tau_R~\right) ~.\ee
and the neutrino masses follow from
\be \label{wnu}
W_\nu=\frac{h_\nu}{2M} (l_L^T C \Delta l_L)_{3_S} \chi_\nu~.\ee
Here, $C$ is the charge conjugation matrix. The subscript $a$ in  $(..)_a$ labels the $A_4$ representation according
to which the  quantity  (...) transforms. The scale $M$ and the flavon vacuum
expectation values generate the effective Yukawa couplings in the model.
The additional symmetry $Z_3$
introduced here serves two purposes. It prevents $\chi_e (\chi_\nu)$
couplings in $W_\nu(W_l)$ at the leading order. It also forbids a singlet
term $(l_L^TC\Delta l_L)_1$ allowed by the $A_4$ symmetry.

The residual symmetries of the leptonic mass matrices are determined from 
the above superpotential by the flavon vacuum expectation values (vev). We shall choose 
these symmetries in accordance with the discussion given in section II.
The symmetry $T_l$
of $M_lM_l^\dagger$ is chosen as $E$ or  $E^2$. This is realized when  vev
$<\chi_e>$ of $\chi_e$ satisfies $E<\chi_e>=<\chi_e>$.This requires equal vev for all
the  three components of $\chi_e$ and leads to a well-known form of $M_l$
considered in many
models based on $A_4$, see for example \cite{He:2006dk}. The charged lepton
mixing is determined in this case by $U_l=U_\omega$ defined in
eq.(\ref{uw}).
Only possible choice within $A_4$ for $S_\nu$ leading to flavour antisymmetry is given by $F$ or its cyclic permutations.
In order to realize this antisymmetry as residual invariance, we impose $S_\nu<\chi_\nu>=-<\chi_\nu>$ with $S_\nu=F$.
This leads to the
configuration $<\chi_1>=0$ and $<\chi_{2,3}>\equiv v_{2,3}$. Here $v_{2,3}$ are complex parameters
in general. It is seen that eq.(\ref{wnu}) then leads to the flavour
antisymmetric mass matrix given in eq.(\ref{text1}) with
$\tan\theta=\frac{|v_2|}{|v_3|}$, $\beta=Arg(v_2v_3^*)$ and appropriately
defined $m_0$. The neutrino mixing matrix in this case is $U_\nu=V_\nu$. The 
resulting mixing pattern is a  special case of
eq.(\ref{ui3}) obtained for $\Delta(12 k^2)$ with $T_l=R(2k,p,q)$ and 
$S_\nu=W(2k,0,k)$. In the present case, 
the residual symmetry $T_l=E^2$ coincides with 
$R(2,0,0)$ and $S_\nu=F=W(2,0,1)$. Thus $U_{i3}$ can be obtained by putting
$p=q=0$ and $\eta=-1$ in eq.(\ref{ui3}). As already discussed, this leads to
a prediction 
$\theta_{23}=\frac{\pi}{4}$ for $\beta=0$ independent of the choice of
$\theta$. The latter can be chosen to give the correct $\theta_{13}$ while
the solar angle and scale remain unpredicted at this stage due to degeneracy in mass.
We now discuss possible perturbations which can generate them

One source of perturbation that we consider comes from the $A_4\times Z_3$
invariant non-leading corrections to $W_\nu$, eq.(\ref{wnu}). These are
given by 
\beqa
\label{nonleading}
W_\nu'&=&\frac{1}{2 M^2}(f_1(l_L^TC\Delta l_L)_{3_s}(\chi_e\chi_e)_{3_s}+
f_2(l_L^TC \Delta l_L)_{1}(\chi_e\chi_e)_{1}+\nonumber \\
&
&f_3(l_L^TC\Delta l_L)_{1^\prime}(\chi_e\chi_e)_{1^{\prime\prime}}
+f_4(l_L^TC \Delta l_L)_
{
1^{\prime\prime}}(\chi_e\chi_e)_{1^{\prime}} )~. \eeqa
Only the first two terms give non-zero contribution when all the
components of
$\chi_e$ acquire equal vev. The neutrino mass matrix including
these correction is given by
\be \label{pert}
\tilde{M}_\nu=m_0\left(
\ba{ccc}
\epsilon_1&c&se^{i\beta}\\
c&\epsilon_1&\epsilon_2\\
se^{i\beta}&\epsilon_2&\epsilon_1\\
\ea\right)~,\ee
where $\epsilon_{1,2}\ll 1$. Eq. (\ref{nonleading}) corrects the leading
order non-zero elements $(\tilde{M}_\nu)_{12}$ and $(\tilde{M}_\nu)_{13}$.
We have absorbed these corrections through redefinition of $\theta,\beta$
and $m_0$. $\epsilon_{1,2}$ are new corrections arising from
eq.(\ref{nonleading}).
Eq.(\ref{nonleading}) does not exhaust all the non-leading corrections. One
could
also write similar corrections to $W_l$ quadratic in $\chi_\nu$. These will
correct  the charged lepton mixing. Similarly, a more elaborate
realistic model leading to the assumed vacuum configuration will also
contain non-leading corrections which may change the leading order vev
assumed here. All these corrections will add more parameters to the model
and we assume their contribution to be small. Here we show that two
parameters $\epsilon_{1,2}$ introduced by eq.(\ref{nonleading}) in
eq.(\ref{pert}) are sufficient to reproduce the neutrino mixing and scales
correctly. They split the degeneracy and 
generate the solar scale and angle correctly. For example, 
\label{set1}
\be
(\theta,\beta,\epsilon_1\epsilon_2)=(0.5904,
-0.1818 -0.0579,  0.1186)~\ee
give the following values of the observables  
\be
\sin^2\theta_{13}\sim
0.024~~,~~\sin^2\theta_{23}\sim 0.455~~,~~\sin^2\theta_{12}\sim
0.307~~,~~\frac{ \Delta_{\odot}}{|\Delta_{atm}|}\sim 0.0317~\ee
which corresponds to (nearly) best fit values obtained for example with a
global fits in \cite{Capozzi:2013csa}.
\section{Summary}
The  bottom up approach of finding discrete symmetry
groups starting with possible symmetries of the residual mass matrices has
been  successfully used  
in last several years to predict leptonic
mixing angles. The residual symmetry assumed in these works leaves the
neutrino mass
matrix $M_\nu$ invariant. We have proposed here a different possibility
in which $M_\nu$ displays antisymmetry as defined in  eq.(3) under a
residual symmetry. Just
like symmetry, the antisymmetry can also come from breaking of some discrete
group $G_f$ as we have illustrated with an example based on $A_4$. The use
of antisymmetry is found to be more predictive than symmetry. 
It is able to restrict
both neutrino masses and mixing angles unlike all the previous works
in this category
which \cite{Lam:2008rs,Lam:2008sh,Lam:2009hn,Lam:2011ag,Toorop:2011jn,deAdelhartToorop:2011re,Altarelli:2012ss,Holthausen:2012wt,
Hu:2012ei,Hernandez:2012ra,Hernandez:2012sk,
Holthausen:2013vba,Holthausen:2013vba,Lavoura:2014kwa,Fonseca:2014lfa,
Hu:2014kca} could predict only mixing angles. Moreover, the antisymmetry
condition by itself is sufficient for 
determining all possible discrete residual antisymmetry operators $S_\nu$
residing in $SU(3)$. This in turn leads to very specific textures of the
neutrino mass matrix  satisfying antisymmetry condition. These are  given
by eqs.(\ref{text1},\ref{text2},\ref{text3},\ref{text4}).

We studied the mixing angle predictions in the specific context of the
groups $\Delta(3N^2)$ and $\Delta(6 N^2)$. The main results obtained are:
\begin{itemize}
\item Only the groups $\Delta(12 k^2) $
with $k=1,2...$ and all $\Delta(6 N^2)$ groups contain the residual
antisymmetry operator. 
\item Of the four possible neutrino mass textures allowed
by antisymmetry,  only texture I having one massless and two degenerate
neutrinos can lead to correct mixing pattern. 
This case provides a very good zeroeth order approximation to reality if
the neutrino mass hierarchy is inverted. 
\item There always exists within these groups 
residual symmetries of $M_lM_l^\dagger$ and $M_\nu$   such that the
atmospheric neutrino
mixing angle is maximal. Correct value of $\theta_{13}$ can be accommodated 
by choosing the unknown angle in eq.(\ref{text1})
appropriately. There also exists other choices of residual symmetries which 
for some groups allow non-maximal values of the
atmospheric neutrino mixing angle as well. 
The results of various cases are summarized in Fig. 1 and Table II.
\item 
The successful texture I still has two free parameters
apart from an overall mass scale. But as we have shown here, predicted
atmospheric mixing angle in many cases is independent of these unknowns.
The reactor angle $\theta_{13}$ depends on it but it is possible to
determine these unknown also by enlarging the residual symmetry and 
we have given an example of a $Z_2\times Z_2$ residual symmetry which can
determine the complete  neutrino mass matrix up to an overall scale  in
terms of group theoretical
parameters alone and have identified $\Delta(600)$ as a possible group which
can give correct $\theta_{13}$ and $\theta_{23}$ with this symmetry.
\end{itemize}
We end this section with a comparison of the present work with 
some earlier relevant works.
\begin{itemize}
\item The texture I, eq.(\ref{text1}) has been extensively studied since
long in the context of $L_e-L_\mu-L_\tau$ global
symmetry
which implies it, see for example \cite{Goh:2002nk} and references therein.
Imposition of this symmetry on the charged lepton mass
matrix
$M_l$ makes it diagonal after  redefinition of $\theta$ appearing in
(\ref{text1}).
Thus the matrix $V_\nu$ as given in eq.(\ref{unu1}) corresponds to the
final mixing matrix which is now not  allowed by the present experimental
constraints. This is not the case here since the $M_lM_l^\dagger$ is
non-trivial with the imposed discrete symmetry.
\item Neutrino mass matrix displaying a specific flavour antisymmetry
namely, $\mu$-$\tau$ antisymmetry was studied in \cite{Grimus:2005jk}.
This antisymmetry was assumed there to hold in the neutrino flavour basis.
In our terminology, this would correspond to study of a  specific
example within the choice (2A) discussed in section IIIB.
The structure of the neutrino mass matrix
and the mixing angle predictions obtained here for this choice agrees with
ref.\cite{Grimus:2005jk} after suitable basis change.
The study presented here is not limited to the  $\mu$-$\tau$
antisymmetry but encompasses all possible antisymmetry operators within
$SU(3)$ and leads to many new phenomenological predictions.
\item The antisymmetry condition, eq.(3), can be converted to the
usually assumed symmetry condition by redefining the operator
$S_\nu\rightarrow i S_\nu$. The new operator
 does not however have unit determinant and would 
belong to a $U(3)$ group. The occurrence of massless state within such group
with condition, eq.(2) was discussed in
\cite{Joshipura:2013pga,Joshipura:2014pqa}. The residual symmetry operators
used there had eigenvalues $(\eta,1,-1)$ (or its permutations) with
$\eta\not=\pm 1$. This coincides with eigenvalues of $iS_\nu$ for texture
IV when $\eta=i$. Only texture IV was considered in \cite{Joshipura:2013pga,Joshipura:2014pqa} and it was shown there that  a large
class of DSG of $U(3)$ imply  $\sin^2\theta_{13}$ to be either   0 or $\frac{1}{3}$
with condition (2). The same conclusion is found to be true
here with eq.(3) and texture IV in case of the group series $\Delta(3 N^2)$ and
$\Delta(6 N^2)$.
\item It is possible to obtain a degenerate pair of neutrinos using
symmetry condition, eq.(2) and DSG of $SU(3)$. This was 
studied for the finite 
von-Dyck groups in \cite{Hernandez:2013vya}  and
for all DSG of $SU(3)$
having three dimensional IR in \cite{Joshipura:2014qaa}. Here, the third
state is not implied to be massless. The case of one massless and two
degenerate neutrinos can follow  from the symmetry condition if DSG of
$U(3)$ are used. This was also
discussed in \cite{Joshipura:2014qaa}. The successful examples found in these two works
are different from here because of the 
difference in the assumed residual symmetries. The cases
studied in the context of DSG of $SU(3)$ and $U(3)$ \cite{Joshipura:2014qaa}
 have texture similar to
 the texture II in the present terminology. It was found there that this
texture can give non-trivial values of $s_{13}^2$, $s_{23}^2$ in several 
$\Delta(6 N^2)$ groups when symmetry condition (2) is used. This does not
happen with the antisymmetry condition in case of texture II as argued here.
On the other hand,
one can obtain correct values for $\theta_{13}$ and $\theta_{23}$ in 
all the $\Delta(3N^2)$ groups with texture I when antisymmetry condition is
employed. Thus 
symmetry and antisymmetry conditions appear complementary to each
other and allow more possibilities for flavour symmetries $G_f$. 
\end{itemize}
\begin{acknowledgments}
It is pleasure to thank Ketan M. Patel for a careful reading of the
manuscript and helpful suggestions. I also thank  Department of Science and
Technology, Government of India for support under the
J. C. Bose National Fellowship programme, grant no. SR/S2/JCB-31/2010.
\end{acknowledgments}
\bibliography{ref-nuflavas.bib}
\end{document}